\documentclass[fleqn,usenatbib]{mnras}

\usepackage{newtxtext,newtxmath}

\usepackage[T1]{fontenc}
\usepackage{booktabs}

\DeclareRobustCommand{\VAN}[3]{#2}
\let\VANthebibliography\thebibliography
\def\thebibliography{\DeclareRobustCommand{\VAN}[3]{##3}\VANthebibliography}

\usepackage{graphicx}	
\usepackage{amsmath}	
\usepackage{adjustbox}
\usepackage{subcaption}
\usepackage{color}

\usepackage{float}

\title[]{Lunar laser ranging analyses of NGLR-1 data with PEP and prospects for MPAc: towards one millimeter range precision}

\author[G. Bargiacchi et al.]{G. Bargiacchi,$^{1}$\thanks{Corresponding author, E-mail: gbargiacchi@lnf.infn.it}
J. Chandler,$^{2,3}$
N. Burzillà,$^{1}$
R. Campagnola,$^{1}$
L.Porcelli,$^{1}$
D. Currie,$^{4}$
J.D. Carpenter,$^{5}$
\newauthor
N. Boersma,$^{5}$
L. Cacciapuoti,$^{5}$
P. Firme,$^{5}$
P. Ruzza,$^{5}$
I. Lebouteiller,$^{5}$
C. Courde,$^{6}$
J. Chabé,$^{6}$
J. Eckl,$^{7}$
\newauthor
S. Merkowitz,$^{8}$
N. Colmenares,$^{8,9}$
R. Reasenberg,$^{2,3}$
I. Shapiro,$^{3}$
S. Dell'Agnello$^{1}$
\\
$^{1}$INFN - Laboratori Nazionali di Frascati (LNF), Via E. Fermi 40, Frascati 00044 Roma, Italy \\
$^{2}$ University of California, San Diego, 9500 Gilman Drive, La Jolla, CA 92093, USA
\\
$^{3}$ Center for Astrophysics, Harvard and Smithsonian, 60 Garden Street, Cambridge, MA 02138, USA
\\
$^{4}$ Department of Physics, University of Maryland, College Park, College Park MD 20742, USA
\\
$^{5}$ European Space Agency (ESA), European Space Research and Technology Centre (ESTEC),  Keplerlaan 1, 2201, AZ Noordwijk, Netherlands
\\
$^{6}$ Université Côte d’Azur, Observatoire de la Côte d’Azur, CNRS, Laboratoire Géoazur, France
\\
$^{7}$ Federal Agency for Cartography and Geodesy, Geodetic Observatory Wettzell, Sackenrieder Strasse 25, Bad Kötzting, Germany 93444
\\
$^{8}$ NASA Goddard Space Flight Center, 8800 Greenbelt Rd., Greenbelt, 20771, MD
\\
$^{9}$ Peraton Inc
}

\date{}

\makeatletter

\def\ps@mnras{%
  \let\@mkboth\@gobbletwo
  \def\@oddhead{}%
  \def\@evenhead{}%
  \def\@oddfoot{}%
  \def\@evenfoot{}%
}
\def\ps@firstpage{%
  \let\@mkboth\@gobbletwo
  \def\@oddhead{}%
  \def\@evenhead{}%
  \def\@oddfoot{}%
  \def\@evenfoot{}%
}
\makeatother

\begin{document}
\pagestyle{empty}
\label{firstpage}
\pagerange{\pageref{firstpage}--\pageref{lastpage}}
\maketitle
\thispagestyle{empty} 

\begin{abstract}
Since 1969, Lunar Laser Ranging (LLR) has provided critical data for investigating Earth–Moon dynamics, lunar internal structure, and tests of General Relativity (GR). However, the measurement precision of historical arrays (such as those from Apollo and Lunokhod missions) is inherently limited by laser pulse broadening induced by lunar librations. To overcome this limit, next-generation single Cube Corner Retroreflectors (CCRs) of large optical diameter have been developed to eliminate pulse dispersion, enabling LLR measurements to advance from centimeter-level precision down to the millimeter scale.
In this work, we present a detailed analysis of LLR observations from NGLR-1 (Next Generation Lunar Retroreflector-1), the first single 10cm-diameter CCR deployed on the lunar surface. Carried out using the Planetary Ephemeris Program (PEP), this study investigates individual range measurements, normal points (NPs), residuals, and observational dispersion collected by the ground stations at Apache Point Observatory Lunar laser ranging station (APOL), Wettzell, and Grasse. The results: demonstrate the success of NGLR-1 in significantly enhancing time of flight measurement precision compared to legacy Apollo and Lunokhod retroreflector arrays; indicate similar good prospects also for ESA's MPAc laser retroreflector (MoonLIGHT with Pointing Actuators) to be launched in early 2027; pave the way for a new era of LLR observations and {enhanced} tests of GR, alternative gravity theories {in general} and improved lunar geophysics and geodesy.
\end{abstract}



\section{Introduction}
\label{intro}

Since 1969, Lunar Laser Ranging (LLR) has been a
vital technique to measure the distance between the Earth and the Moon by using the time of flight of laser pulses from Earth that are reflected from retroreflectors on the lunar surface. This method indeed offers crucial information about the dynamics of the Earth-Moon system, the Moon’s internal structure, lunar ephemerides, and important tests of General Relativity (GR) \citep{2012CQGra..29r4005M}. LLR has historically relied on the retroreflector arrays of Apollo 11, 14, and 15 and Lunokhod 1 and 2, which contributed a negligible fraction of the instrumental LLR error
budget. However, over the past five decades, the ground station technology has significantly improved, reducing the measurement errors and currently the instrumental LLR error budget is dominated by the spread of the returning laser pulse {originating in} the large size of the arrays due to the impact of lunar librations caused
by the Moon’s orbital eccentricity and inclination \citep{2012P&SS...74..276M,2017AdSpR..60.1300C,porcelli2021next}. Indeed, this effect is proportional to the dimension of the retroreflector array and it results in a broadening of the laser return pulse that intrinsically limits the precision of the Earth-Moon distance measurement. This means that the spread (i.e.{,} the standard deviation) of the reflected pulse is a measure of the precision of the LLR observation. \\
In this scenario, next-generation single lunar Cube Corner Retroreflectors (CCRs) with large optical diameter allow us to overcome this problem \citep{1970Sci...167..458A,alley1970preliminary,1973Sci...182..229B,currie2011lunar}, as their performance is unaffected by the time spread. As a matter of fact, the pulses reflected by single large-diameter CCRs return separately and can be clearly distinguished in time and space by LLR Earth stations. As a consequence, they are capable of fully exploiting the current LLR station capabilities to reach a precision below the centimeter threshold down to the desired millimeter level, providing also higher data collection rates.
Such advancement will have a significant impact, enabling more accurate ephemerides, refined tests of GR and of other theories of gravity, and improved
knowledge of the properties of the deep lunar interior.\\
The first next-generation large-diameter CCR has already landed on the Moon: it is the Next Generation Lunar Retroreflector-1 (NGLR-1)\footnote{\url{https://www.physics.umd.edu/nglr/}.} \citep{currie2011lunar}. NGLR-1 is a single CCR with an optical diameter of 10 cm developed by the University of Maryland in collaboration with the Istituto Nazionale di Fisica Nucleare - Laboratori di Frascati (INFN-LNF). It was deployed on the Moon on the 2nd of March 2025 in Mare Crisium by Firefly Aerospace’s Blue Ghost lunar lander {\citep{2026LPICo3160.1629B,2026Meas..28322094Y}}. Since the first day after landing, NGLR-1 has been observed by the Earth stations of Grasse in France, Wettzell in Germany, and Apache Point Observatory Lunar laser ranging (APOL) station in the USA, formerly known as the Apache Point Observatory Lunar Laser-ranging Operation (APOLLO). These stations are collecting normal points\footnote{A normal point is a compressed measurement of the round-trip time of flight between a ground station and the retroreflectors on the Moon, obtained by averaging a series of raw observations (i.e.{,} single photons or individual range measurements) collected over a specific time interval {typically lasting 5 to 15 minutes}. {To compress data, observatories isolate true photon returns from background noise during the time window; these valid returns are then mathematically mapped to a single central time and averaged to smooth out random timing variations. This process reduces data volume while increasing measurement precision}.} (NPs) from NGLR-1, and we also expect measurements by the station located in Matera, Italy. Furthermore, an upgrade of the laser and detector systems of the ground stations is planned to enhance the precision of the measurements and provide a polarization angle in agreement with the one designed for NGLR-1. Finally, other LLR stations are expected to acquire data from NGLR-1 in the future. In addition, new next-generation CCRs are expected to land on the Moon. In particular, the successors of NGLR-1 are NGLR-2 and NGLR-3 and the one developed by INFN-LNF, which is MoonLIGHT+MPAc (Moon Laser Instrumentation for General relativity/geophysics High-accuracy Tests integrated with the MoonLIGHT Pointing Actuator) \citep{2024AstTI...1..295H,muccino2025moonlight}.\\
The main objective of NGLR-1 is to significantly improve the precision of the LLR measurements. In this regard, the data obtained by the operating ground stations provide important information on the ranging dispersion and the precision of the measurements. In this framework, we here present an analysis of the NGLR-1 observations that deeply investigates the individual range measurements, NPs, residuals, and spread of the observations to NGLR-1. This study is performed with the Planetary Ephemeris Program. This work clearly shows the success of NGLR-1 in enhancing the precision of the time of flight measurement compared to the "old-generation" retroreflector arrays, paving the way to a new era of LLR observations that allow for a breakthrough in {some of the tests} of GR and in general gravity theories.\\
The paper is organized as follows: Sect. \ref{data} introduces the data sets used in this analysis, Sect. \ref{methods} details the use of the Planetary Ephemeris Program, its specifics, and our methodology for dealing with the LLR data, Sect. \ref{results} outlines the results of the analyses, and Sect. \ref{mpac} describes the MoonLIGHT+MPAc payload and its expected impact on the LLR analyses performed with PEP. A summary of conclusions and future prospects is provided in Sect. \ref{conclusions}.

\section{LLR data sets}
\label{data}

In this work, we use three main separate LLR data sets, which are those acquired from APOL, Wettzell, and Grasse stations on NGLR-1. The data from APOL and Wettzell stations are publicly available on the online repository of the Crustal Dynamics Data Information System (CDDIS\footnote{\url{https://cddis.nasa.gov/archive/slr/data}}) and EUROLAS Data Center (EDC\footnote{\url{https://edc.dgfi.tum.de/en}}), while the Grasse data set {was exchanged privately between the parties and is not in the public domain due to an internal policy decision by the ground station}.\\
In more detail, the APOL data set consists of 46 NPs obtained in the period between the 20th of March 2025 and the 9th of April 2026, at a wavelength of 532 nm. 
The NPs from the Wettzell station are two at a wavelength of 1064 nm. In particular, they consist of 37 and 32 individual range measurements (i.e.{,} single photon measurements and not NPs) acquired on the 4th of March 2025, the second day after the deployment of NGLR-1 on the Moon.
The Grasse data are composed of 80 NPs, of which 11 NPs {are} at 532 nm and 69 NPs at 1064 nm, acquired from the 3rd of March 2025 to the 30th of April 2025. More precisely, the NPs at 532 nm {all} belong to the same day of observation, that is the 4th of March 2025. For both the data sets, the individual range measurement files are also available.
We need to treat the LLR data from the APOL station in a different way compared to the data from Wettzell and Grasse since the APOL observations are released in the repository only in the format of NPs (without the corresponding individual range measurements). This {difference} is detailed in Sect. \ref{results}.

\section{The Planetary Ephemeris Program and our methods}
\label{methods}

In this section, we detail our methodology as follows: in Sect. \ref{pep} we describe the Planetary Ephemeris Program (PEP) along with its specifics and setup, in Sect. \ref{crdtomini} we delineate the necessity of converting LLR data from the format of online repositories into a PEP-readable format and how to perform the conversion, and in Sect. \ref{detrending} we outline the detrending procedure and its importance to isolate and analyze real signals removing temporal trends.

\subsection{The Planetary Ephemeris Program}
\label{pep}

PEP is an open-source software\footnote{\url{https://gitlab.com/jbattat/}} developed since the 1960s {at first at the Massachusetts Institute of Technology (MIT) and then at the} Harvard-Smithsonian Center for Astrophysics (CfA) in Fortran language \citep{1988PhRvL..61.2643S,2015PhDT........65J,2021AJ....162...78C}. It uses LLR data for tests of GR and beyond {(such as geometric and metric-affine extensions, vector, tensor and scalar-tensor theories)} by comparison of GR predictions with observations in the frame of the Parametrized Post-Newtonian (PPN) formalism \citep{11114645, bargiacchi_2026_20381367}. In this scenario, PEP can improve the accuracy of the constraints on the weak and strong equivalence principles, the variation with time of the gravitational constant, the geodetic precession, and {also} put constraints on theories beyond GR \citep{2026PhRvD.114b4009B}, like spacetime torsion \citep{2011PhRvD..83j4008M}, f(R) gravity \citep{2019IJGMM..1650008C}, non-minimally coupled gravity \citep{2022PhRvD.105d4048M,2024PhRvD.109l4013M,2026PhRvD.113h4017M} and so on. To this aim, PEP is based on a very detailed and accurate mathematical model of the whole Solar System that includes a huge number of adjustable parameters related to planets, asteroids, {other} celestial bodies and, in general, to the physics of the Universe. However, for this work we have focused only on the Earth-Moon-Sun system to deal with LLR data.\\
PEP follows three main steps: i) the integration of the equations of motion to generate the ephemeris, ii) the computation of the residuals (O-C) of the time of flight; these are the differences between the observed (O) time of flight and the one calculated (C) from the physical model embedded in PEP, and iii) the estimation of the parameters of interest and their associated uncertainties {from these "(O-C)"s}. {For} the scope of this paper, we are mainly interested in the second step, so we consider that we have already generated the ephemeris and we work on the computation and plot of the residuals for the LLR data sets described in Sect. \ref{data}. We here notice that the time unit used by PEP is the Julian {Day} (JD), which can be converted to calendar date by using online converters\footnote{We recommend the converter at \url{https://aa.usno.navy.mil/data/JulianDate}.}.

\subsection{The conversion of the data format}
\label{crdtomini}

LLR data from 2012 are publicly available in the online databases of CDDIS and EDC in a common format, the Consolidated laser Ranging Data format (CRD\footnote{A detailed description of this format can be found at \url{https://ilrs.gsfc.nasa.gov/docs/2022/crd_v2.01e3.pdf}.}).
However, PEP requires files used to generate LLR data libraries in a different format, the one historically used for the APOL data\footnote{Description of the format and corresponding APOL data files until the end of year 2020 at \url{https://tmurphy.physics.ucsd.edu/apollo/norm_pts.html}.}. For this reason, we have identified a one-to-one correspondence between the fields in one format and the fields in the other format, and we have produced a Python code that performs the conversion from CRD into the PEP-readable format, handles different Earth stations and works both for normal point and individual range measurement files.
This conversion is a crucial point to introduce in PEP new and more recent LLR data that are not implemented in the publicly available version of PEP.

\subsection{The detrending process}
\label{detrending}

Once the NGLR-1 data have been converted into the required format, we can process them with PEP and compute the residuals. However, we need a further step to properly analyze the residuals, that is the detrending process. Detrending the residuals in LLR is a crucial post-processing step used to remove systematic, time-dependent errors from raw time of flight measurements, isolating the true physical signal (the Earth-Moon distance) from the noise. The process enables the achievement of millimeter-level precision by removing temporal trends within a specific observational session. The main goal is to remove temporal trends that may arise during a single, long observation session due to atmospheric changes or minor modeling inaccuracies, 
allowing for cleaner and more accurate data
(see also \citealt{2020EP&S...72..113M}). The detrend is applied by calculating a fit of the residuals over the course of a single session and then removing the obtained trend to eliminate systematic slopes. The crucial point is that the degree of the polynomial detrending function must be chosen accurately based on statistical criteria since the choice of the degree of the polynomial significantly modifies the standard deviation of the detrended residual distribution; eventually, a wrong choice of the polynomial degree (and mainly high-order degrees) could alter the real signal and invalidate the analysis.\\

\begin{table}
    \centering
    \begin{tabular}{ccccc}
        \hline
        Degree & RMS (ps) & AIC & Durbin-Watson & Reduced $\chi^2$ \\
        \hline
        1 & 2302.53 & -1467.81 & 0.047 & 3313.52 \\
        {2} & {40.08} & {-1765.57} & {2.118} & {1.00} \\
        3 & 39.24 & -1765.14 & 2.192 & 0.96 \\
        4 & 39.22 & -1763.17 & 2.196 & 0.96 \\
        5 & 39.14 & -1761.33 & 2.215 & 0.96 \\
        \hline
    \end{tabular}
     \caption{Results of statistical tests performed for the polynomial detrending for the 37 Wettzell LLR points across different polynomial degrees.}
    \label{tab:detrending_analysis}
\end{table}

\begin{figure}
\centering
\begin{subfigure}[b]{0.75\linewidth}
\includegraphics[width=0.8\linewidth]{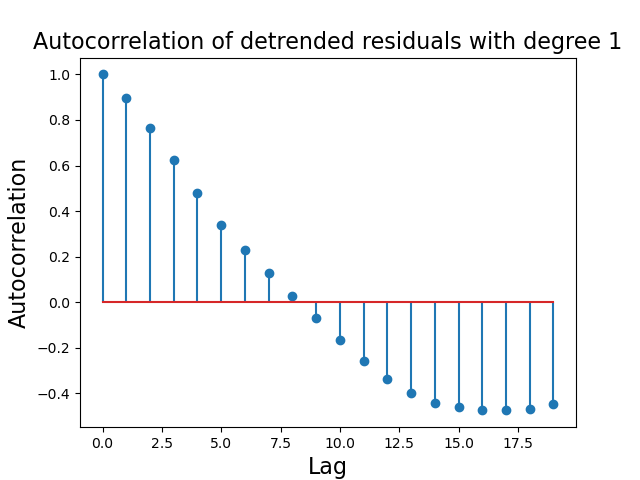}
\caption{Autocorrelation function plot for a 1-degree polynomial detrending.}
\end{subfigure}
\begin{subfigure}[b]{0.75\linewidth}
\includegraphics[width=0.8\linewidth]{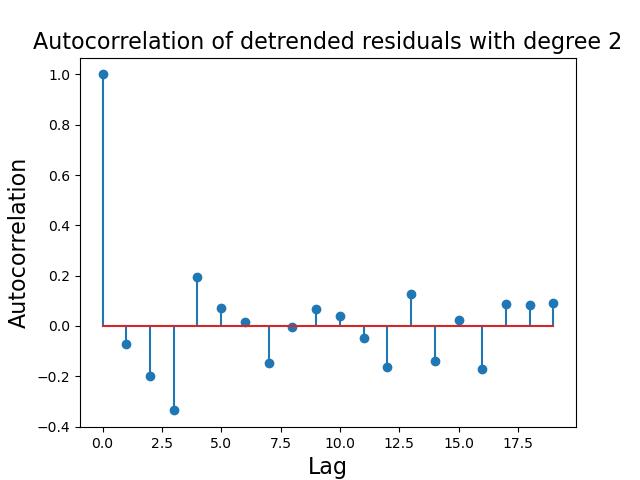}
\caption{Autocorrelation function plot for a 2-degree polynomial detrending.}
\end{subfigure}
\begin{subfigure}[b]{0.75\linewidth}
\includegraphics[width=0.8\linewidth]{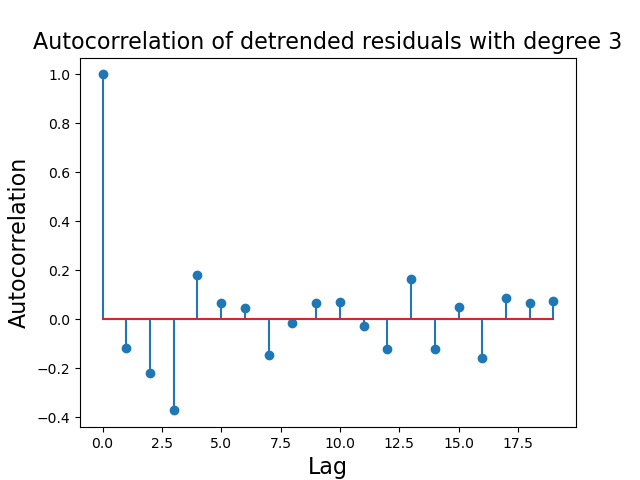}
\caption{Autocorrelation function plot for a 3-degree polynomial detrending.}
\end{subfigure}
\begin{subfigure}[b]{0.75\linewidth}
\includegraphics[width=0.8\linewidth]{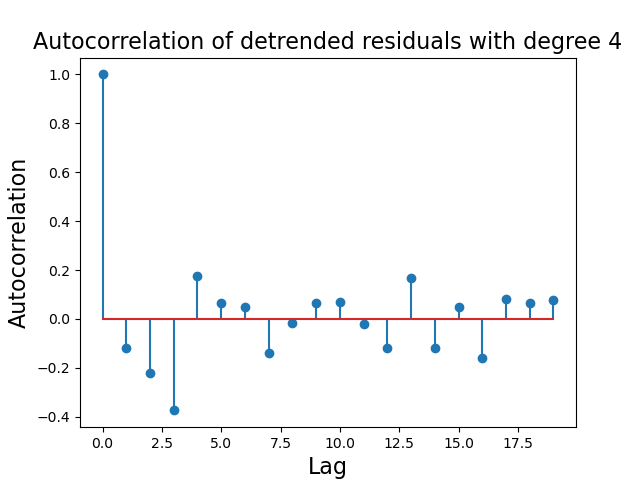}
\caption{Autocorrelation function plot for a 4-degree polynomial detrending.}
\end{subfigure}
\begin{subfigure}[b]{0.75\linewidth}
\includegraphics[width=0.8\linewidth]{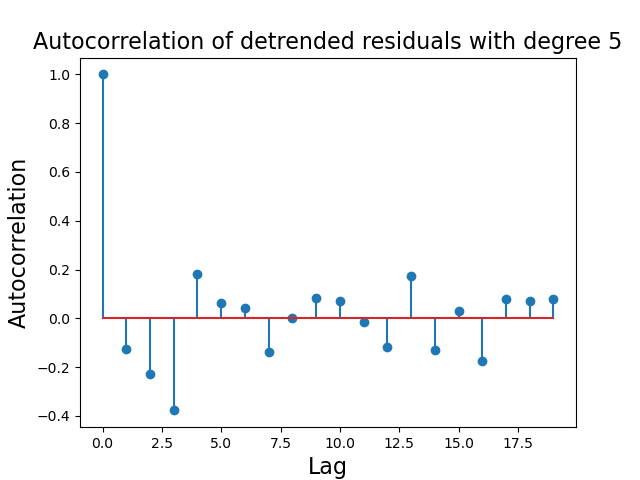}
\caption{Autocorrelation function plot for a 5-degree polynomial detrending.}
\end{subfigure}
\caption{Plots of autocorrelation function across different polynomial degrees for the 37 Wettzell LLR points.}
\label{fig:ACF_plots}
\end{figure}

\begin{figure}
\centering
\begin{subfigure}[b]{\linewidth}
\includegraphics[width=\linewidth]{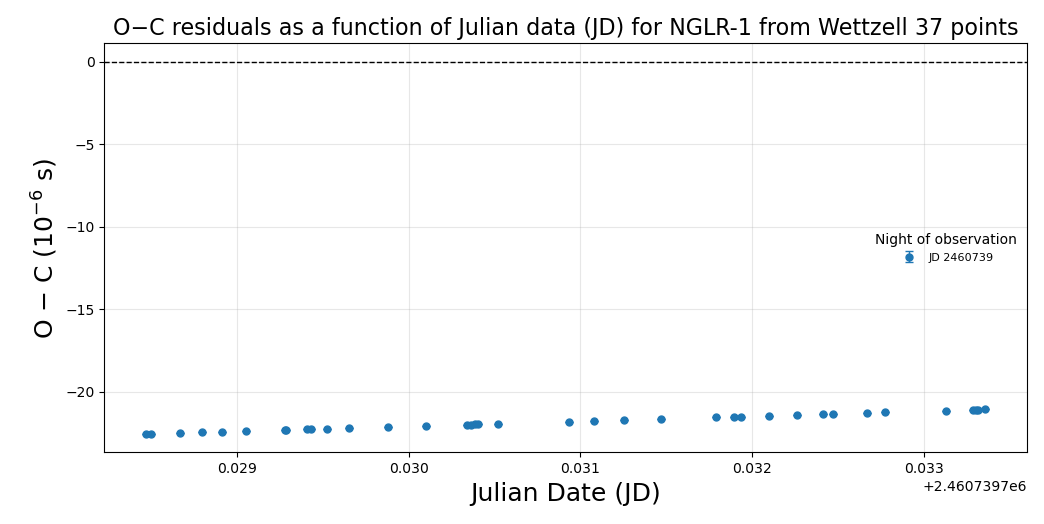}
\caption{Residuals as a function of time before detrending}
\end{subfigure}
\begin{subfigure}[b]{\linewidth}
\includegraphics[width=\linewidth]{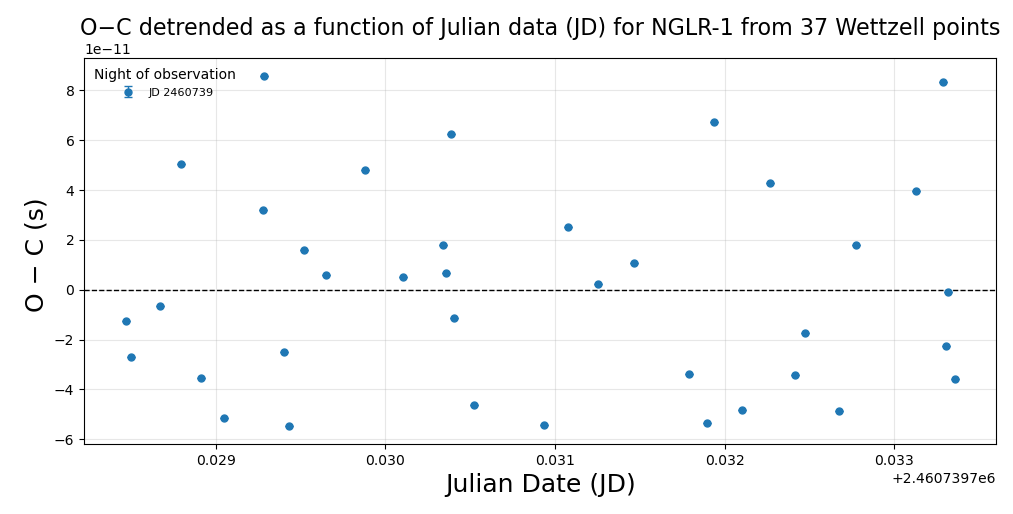}
\caption{Residuals as a function of time after detrending}
\end{subfigure}
\begin{subfigure}[b]{\linewidth}
\includegraphics[width=\linewidth]{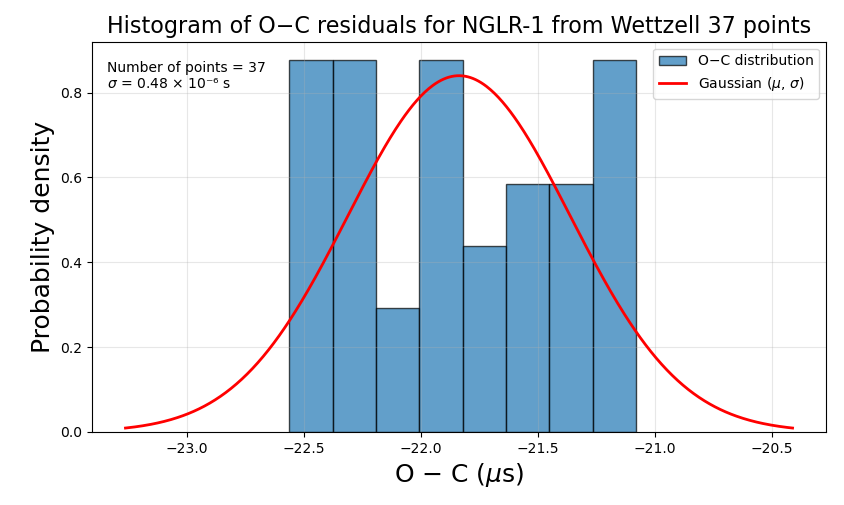}
\caption{Histogram of residuals before detrending}
\end{subfigure}
\begin{subfigure}[b]{\linewidth}
\includegraphics[width=\linewidth]{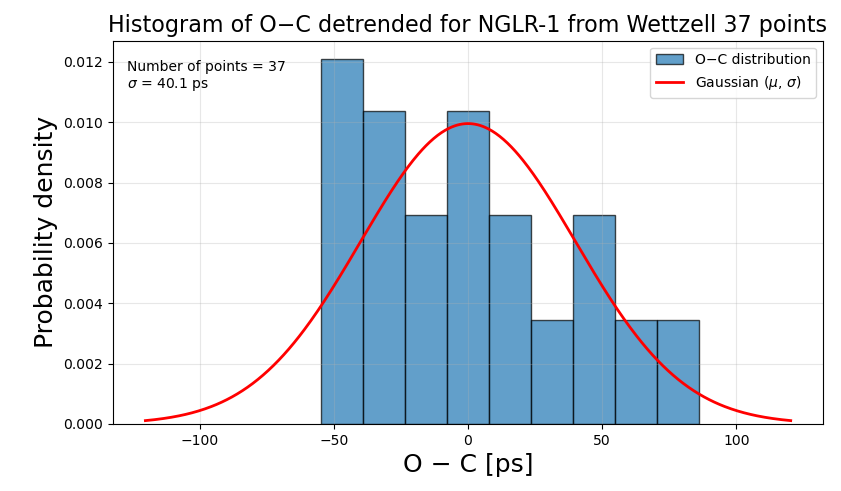}
\caption{Histogram of residuals after detrending}
\end{subfigure}
\caption{Residuals O-C as a function of the time (top panels) and histograms of the residuals (bottom panels) before and after the detrending, respectively on the left and the right column, for the NGLR-1 37 individual range measurements from the Wettzell station. The bottom panels show the data in blue and the associated Gaussian curve in red with the standard deviation $\sigma$ displayed in the {upper} left corner of the figure.}
\label{fig:detrending}
\end{figure}

To face this, we have generated a Python code that performs some independent statistical tests on the raw residuals to identify the proper detrending polynomial function. The tests are as follows {and an example of the statistical results is provided in Figure \ref{fig:ACF_plots} and Table \ref{tab:detrending_analysis}}:
\begin{itemize}
    \item Root Mean Square (RMS) {of residuals: it evaluates the dispersion between the observed residuals and the prediction generated by a polynomial model of a certain degree. As the polynomial degree increases, the model gains flexibility, which naturally decreases the RMS value. While a lower RMS indicates a tighter fit to the data, continually increasing the polynomial degree introduces the risk of overfitting. In this regime, the polynomial begins to fit noise and experimental errors rather than the underlying pattern. As a consequence, defining the most suitable polynomial requires finding the optimal trade-off between minimizing the RMS and preventing overfitting};
    \item Akaike Information Criterion (AIC): a model selection tool that balances goodness-of-fit against model complexity {to determine the optimal polynomial degree}, that is the one that minimizes the AIC value {(for more details see \citealt{akaike1974new,burnham2002model}). An overly high polynomial degree risks fitting random noise or absorbing real physical/geodetic signals of interest. AIC provides an objective statistical criterion to strip away instrumental trends while preserving the integrity of the LLR data. Its application to LLR residuals can reduce subjective model selection, however, AIC alone cannot distinguish instrumental artifacts from genuine physical signals. Preservation of the physical information therefore requires independent validation;}
    \item Autocorrelation Function (ACF): it checks whether residuals behave like white noise; it measures how a signal correlates with a delayed, time-shifted version of itself (see also Figure \ref{fig:ACF_plots}). Ideally, the ACF (on the y-axis) should be 1 at lag 0 (the quantity on the x-axis) and near 0 for all other lags, where the lag is the amount of time shifted. {In particular, in finite data sets, ACF values for non-zero lags are expected to vary slightly around zero rather than vanishing completely, provided they remain within statistical confidence intervals};
    \item Durbin-Watson statistic {\citep{durbin1950testing,durbin1992testing}}: a test for first-order autocorrelation; a value near 2 indicates independent residuals (no trend left), while values below 2 suggest positive correlation and viceversa. The Durbin-Watson method output also depends on the number of points: the acceptable interval around 2 is given by $2 \pm 2/\sqrt{N}$, where $N$ is the number of points considered. {This tolerance interval defines the region where the null hypothesis cannot be rejected; falling within this range confirms that the polynomial detrending model has successfully stripped out systematic trends, leaving pure random noise. Indeed, on the one hand, a positive autocorrelation (with values of the statistic lower than 2) indicates an uncaptured systematic drift or trend (i.e., underfitting or using a polynomial degree that is too low). On the other hand, a negative autocorrelation (with values of the statistic greater than 2) indicates over-correcting};
    \item Reduced $\chi^2$ ($\chi^2_{\text{red}}$): it compares the residual variance against the known experimental uncertainties {according to the formula $\chi^2_{\text{red}} = \frac{1}{N - k} \sum_{i=1}^{n} \left( \frac{(O-C)_i}{\sigma_i} \right)^2$, where $N$ is the number of data points, $k$ is the number of estimated polynomial parameters, and $(O-C)_i$ and $\sigma_i$ are the residual and experimental uncertainty of the observation $i$, respectively. The interpretation of the resulting value is as follows:
    \begin{itemize}
    \item {$\chi^2_{\text{red}} \approx 1$:} the detrending model is fully consistent with the expected noise level and experimental error bars;
    \item {$\chi^2_{\text{red}} \gg 1$:} suggests unmodeled systematic trends (underfitting) or underestimated experimental errors;
    \item {$\chi^2_{\text{red}} \ll 1$:} points to potential overfitting or overestimated measurement uncertainties.
\end{itemize}}
\end{itemize}
{The combination and cross-validation of these tests provide a robust basis for identifying the proper polynomial degree required to detrend the raw residuals.}\\
{To provide a clearer description of the above points and in general the statistical analysis performed to identify the proper polynomial degree for the detrending, we here present the results of each statistical test described above for the specific case of reference of the 37 points from the Wettzell station on NGLR-1 (see Sect. \ref{data}). For conciseness, results are presented for this representative case only, even though nearly  identical analyses have been conducted for all the data sets studied. Table \ref{tab:detrending_analysis} reports the results of RMS, AIC, Durbin-Watson, and reduced $\chi^2$ tests across different polynomial degrees, from degree 1 to degree 5. The visual results of the ACF test for degrees 1-to-5 are shown in the panels of Figure \ref{fig:ACF_plots}. Based on these results, we can draw the following conclusions:
\begin{itemize}
    \item RMS: a massive quality jump occurs when going from degree 1 ($2302.53\text{ ps}$) to degree 2 ($40.08\text{ ps}$) and subsequent degrees yield negligible improvements, in the order of fractions of a picosecond ($39.24 \to 39.14\text{ ps}$), which do not justify the increased polynomial complexity;
    \item AIC: the value reaches its minimum at degree 2 ($-1765.57$), while, as the degree increases, the AIC increases, indicating that the penalty for additional parameters outweighs the likelihood gain, signaling that higher degrees are starting to overfit;
    \item Durbin-Watson and ACF: the ideal Durbin-Watson expected value falls in the range $2 \pm 2/\sqrt{37} = 2 \pm 0.329$, thus only the degree 1 can be rejected, while degrees from 2 to 5 provide results within the confidence interval. However, the degree 2 value is $2.118$, the one closest to $2$, indicating that the residuals are free from significant autocorrelation, while higher degrees risk negative correlation and thus an over-correction. This is strongly corroborated by the ACF plots: while degree 1 (panel (a) of Figure \ref{fig:ACF_plots}) exhibits a strong, slowly decaying autocorrelation across the lags, degree 2 (panel (b) of Figure \ref{fig:ACF_plots}) successfully removes this pattern, with autocorrelation values dropping sharply from the expected value of 1 at lag 0 and fluctuating randomly around zero within acceptable bounds at higher lags. The ACF plots for degrees 3 to 5 (panels (c), (d), and (e) of Figure \ref{fig:ACF_plots}) remain essentially identical to the case of degree 2, confirming that higher polynomial orders yield no further improvement in residual independence;
    \item $\chi^2_{\text{red}}$: an extreme jump occurs when going from degree 1 ($3313.52$) to degree 2, that is exactly $1.00$, representing the theoretical ideal value that indicates a perfect agreement between the model and the expected data variance. From degree 3 onwards, $\chi^2_{\text{red}}$ drops to $0.96$, which is a typical symptom of a slightly overfitted model.
\end{itemize}
In conclusion, considering the trade-off between model fit, residual independence, and physical consistency, a second-degree polynomial represents the optimal choice for detrending the LLR residuals of this data set.}\\
We can now compare the residuals and their distribution before and after the detrending procedure. This comparison is visually reported in Figure \ref{fig:detrending} for the case of reference above studied of the 37 points from Wettzell on NGLR-1, where a 2-degree polynomial detrending has been applied. {For visual clarity, we show here only this case as an example since we have checked that} the same comparison analysis provides the same results on all data sets studied in this work. Comparing panels (a) and (b), it is clear that, on the one hand, the raw residuals {(panel (a))} present a strong trend with time{, which is almost linear, due to the time span of the observation}, {occupy the region} under the zero {dashed} horizontal line, and are of the units of $10^{-5}$ s. On the other hand, the residuals post-detrending {(panel (b))} are randomly scattered along the zero line without any temporal trend and the units of the O-C have reduced to the order of $10^{-11}$ s. In parallel, the standard deviation significantly decreases from $4.8 \cdot 10^{-7}$ s in panel (c) to $4.0 \cdot 10^{-11}$ s in panel (d), where $\sigma$ is computed for the two-way range (see Sect. \ref{results} for details). 
This analysis clearly shows the crucial role of the detrending process for a proper analysis of the true signal.

\section{Results and Discussion}
\label{results}

We here outline our results from the LLR data sets under study in this analysis. In particular, we provide the description and plots of the residuals O-C as a function of time, expressed in JDs, and the distribution of the residuals along with their associated standard deviation. The estimate of the standard deviation is indeed of particular importance, as already stated in Sect. \ref{intro}, since {the standard deviation is a quantitative measure of} the precision of the Earth-Moon distance that {can be reached based on} the capabilities {and limitations} of both the retroreflector and the station of observation.

\subsection{Results from the APOL station}
\label{APOLLOresults}

\begin{figure}
\centering
\includegraphics[width=\linewidth]{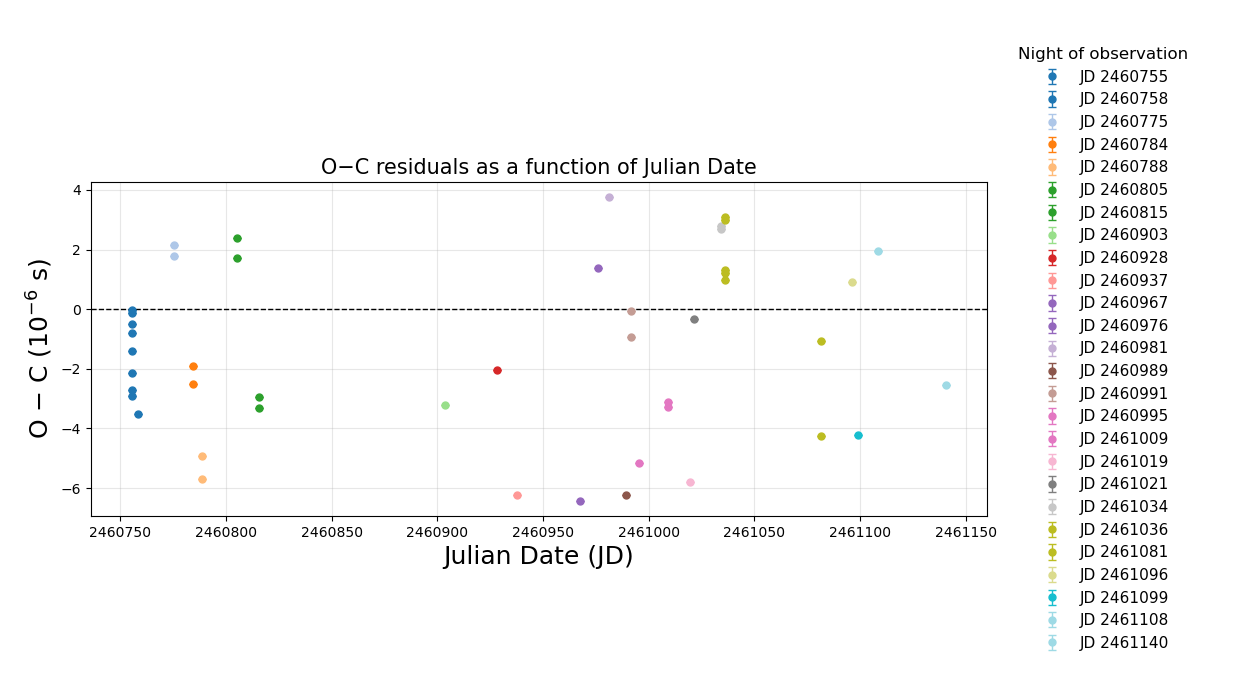}
\caption{Residuals O-C without the detrending for {the} NGLR-1 46 NPs from the APOL station.}
\label{fig:residuals_APOLLO_withoutdetrending}
\end{figure}

As anticipated in Sect. \ref{data}, we deal with the LLR data from the APOL station in a dedicated way since the observations cover a long time range of about one year and are only released as NPs without the corresponding files for the individual range measurements. This implies that we cannot apply the detrending,
 as instead in the case of the Wettzell and Grasse data. As a consequence, we show the analyses without detrending for the whole data set. This analysis does not allow us to compare with the results from the Wettzell and Grasse data, which are described in Sects. \ref{Wettzellresults} and \ref{Grasseresults}, respectively, but this study is beneficial to our work since it clearly provides an inspection of how non-detrended residuals{, compared to detrended residuals,} behave as a function of time and in their probability density.\\
Starting from the total of 46 NPs, we show in Figure \ref{fig:residuals_APOLLO_withoutdetrending} the residuals as a function of the JD, color-coded by the day of observation as shown in the legend.

\begin{figure}
\centering
\includegraphics[width=\linewidth]{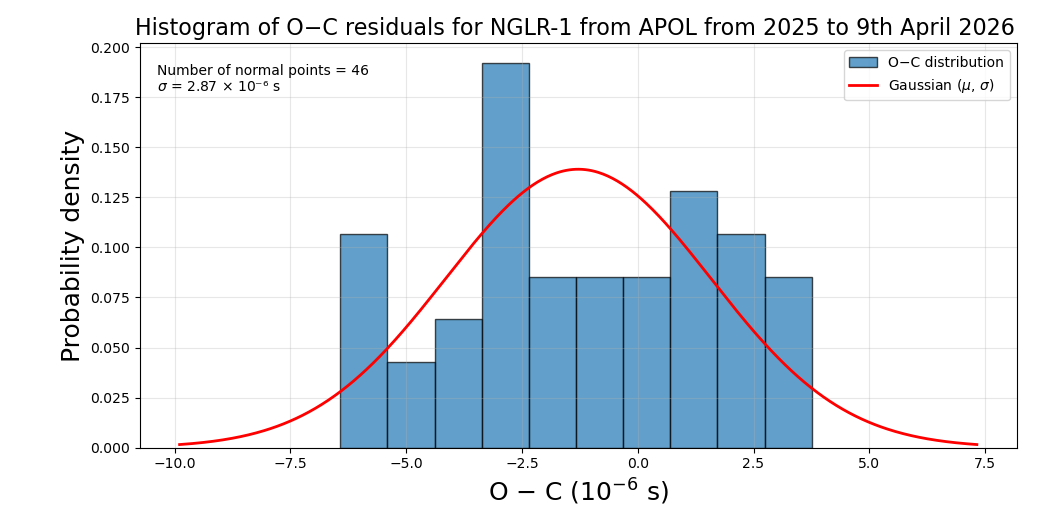}
\caption{Histogram of the residuals O-C, without the detrending for {the} NGLR-1 46 NPs from the APOL station. The blue data illustrate the observations made by the APOL station, while the red curve is the Gaussian curve with the standard deviation associated with the {same} data.}
\label{fig:hist_APOLLO_withoutdetrending}
\end{figure}

Correspondingly, Figure \ref{fig:hist_APOLLO_withoutdetrending} shows the histogram of the residuals of the 46 NPs.
We can notice that the order of magnitude of the O-C in Figures \ref{fig:residuals_APOLLO_withoutdetrending} and \ref{fig:hist_APOLLO_withoutdetrending} is $10^{-6}$ s and that {points acquired during the same day of observation (those with the same color in Figure \ref{fig:residuals_APOLLO_withoutdetrending}) present a visible trend with time}, as expected since the data are not detrended (see Sect. \ref{detrending} for more details). Moreover, the standard deviation $\sigma$ of the distribution is $\sim 3 \mu$s, as reported in Figure \ref{fig:hist_APOLLO_withoutdetrending}.

In the next sections, we perform the same analysis on the Wettzell and Grasse data, showing the significant difference between NPs spanning months of observations and individual range measurements on the same day from the point of view of both the magnitude of the residuals and their standard deviation.

\subsection{Results from the Wettzell station}
\label{Wettzellresults}

\begin{figure}
\centering
\begin{subfigure}[b]{\linewidth}
\includegraphics[width=\linewidth]{O-C_Wtzl_37points_detrended.png}
\caption{For 37 individual range measurements}
\end{subfigure}
\begin{subfigure}[b]{\linewidth}
\includegraphics[width=\linewidth]{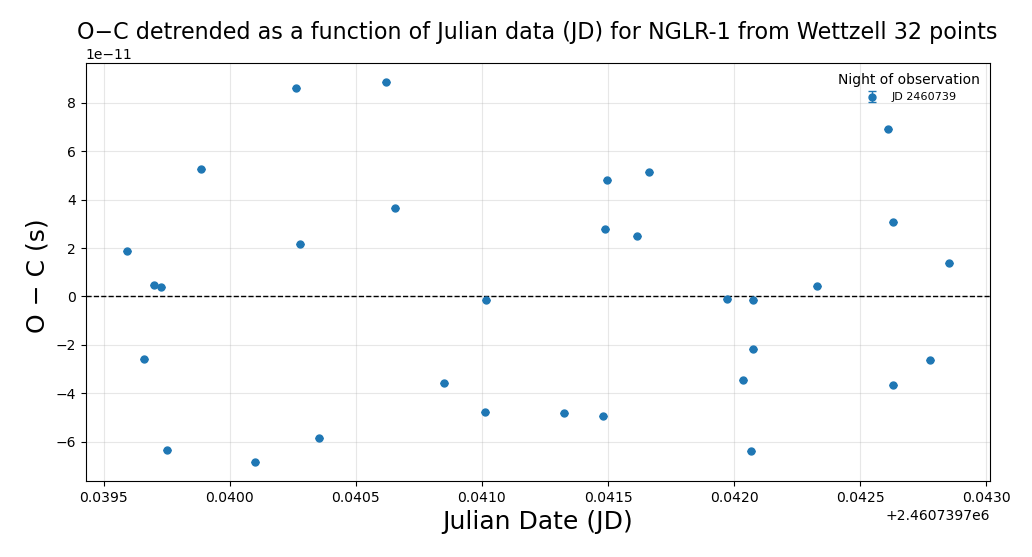}
\caption{For 32 individual range measurements}
\end{subfigure}
\begin{subfigure}[b]{\linewidth}
\includegraphics[width=\linewidth]{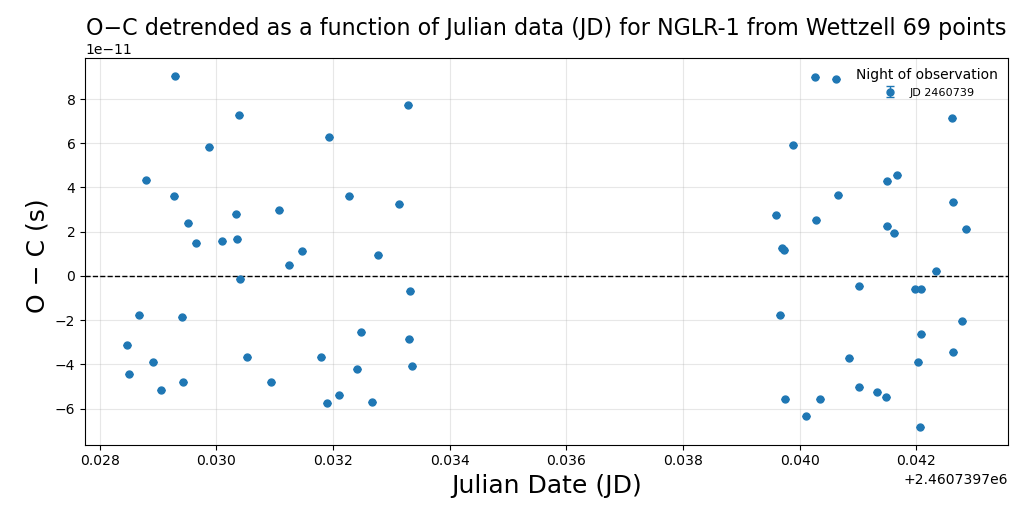}
\caption{For the combined 69 individual range measurements}
\end{subfigure}
\caption{Residuals O-C after the detrending for NGLR-1 individual range measurements from the Wettzell station. Panels (a), (b), and (c) show the case of the 37, 32, and the total of 69 points, respectively.}
\label{fig:residuals_Wtzl}
\end{figure}

Since the Wettzell data set is divided into two NPs acquired on the same day with 37 and 32 individual range measurements, respectively (see Sect. \ref{data}), we have performed our analysis both on the separated and the combined data.
As a first step, we have computed the "raw" (i.e. before detrending) residuals with PEP. Then, we have identified the proper detrending function following the statistical analysis described in Sect. \ref{detrending}. As a result, we have detrended the two separate data sets with a 2-degree polynomial function.
The residuals obtained after the detrending for the three cases (i.e.{,} two separated data sets and the combined one) are shown in the panels of Figure \ref{fig:residuals_Wtzl}. The points are color-coded based on the JD of observation, which, in this case, is the same one for all the measurements.
We can notice the first relevant difference compared to the analysis of the APOL data set studied in the previous section: as anticipated in Sect. \ref{detrending}, the detrended residuals are randomly scattered about the horizontal zero dashed line without any temporal trend, and their magnitude is significantly reduced to the order of $10^{-11}$ s (from $10^{-5}$ s for the residuals before the detrending).

\begin{figure}
\centering
\begin{subfigure}[b]{\linewidth}
\includegraphics[width=\linewidth]{hist_Wtzl_37points_detrended.png}
\caption{For 37 individual range measurements}
\end{subfigure}
\begin{subfigure}[b]{\linewidth}
\includegraphics[width=\linewidth]{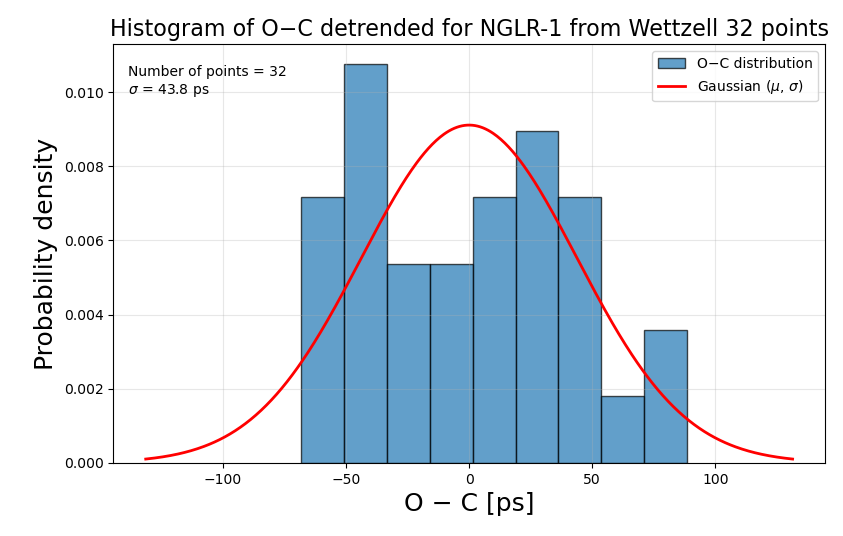}
\caption{For 32 individual range measurements}
\end{subfigure}
\begin{subfigure}[b]{\linewidth}
\includegraphics[width=\linewidth]{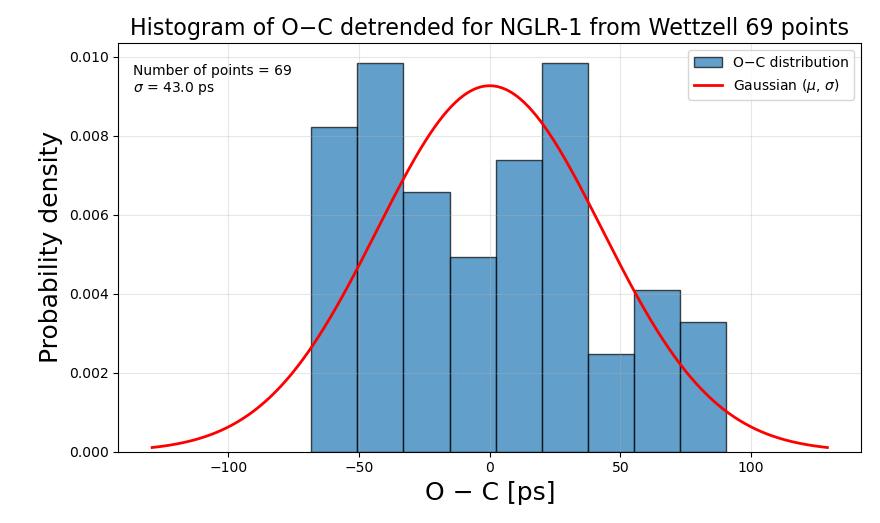}
\caption{For the combined 69 individual range measurements}
\end{subfigure}
\caption{Histograms of the residuals O-C after the detrending for NGLR-1 individual range measurements from the Wettzell station. Panels (a), (b), and (c) show the cases of the 37, 32, and the total of 69 points, respectively. The blue data illustrate the observations made by the Wettzell station on the 4th of March 2025, while the red curve is the Gaussian curve with the standard deviation associated with the {same} data.}
\label{fig:hist_Wtzl}
\end{figure}

After the inspection of the residuals, we plot the three distributions of the O-C, and we superimpose the Gaussian distribution with the corresponding standard deviation{, that is computed as $\sigma = \sqrt{\frac{1}{N} \sum_{i=1}^{N} (x_i - \bar{x})^2}$, where $N$ is the number of points, $x_i$ the single residual of $i$-observation and $\bar{x}$ the average of residuals}. Figure \ref{fig:hist_Wtzl} reports the results for the different data sets, as described in the caption of each panel. The legend in each panel also describes the data shown and displays the standard deviation, $\sigma$. We can notice that the standard deviations are of the order of 40 ps in all the cases studied. We here pinpoint that, for visual clarity, the $\sigma$ reported in the plots is computed as the standard deviation associated with the residual distribution presented{, as shown in the previous formula}. This means that $\sigma$ refers to the residuals of the round trip time of flight (i.e., from the Earth to the Moon and back), since this is the measured quantity and the one reported in the data files. However, we are interested in the standard deviation associated with the one-way measurements, i.e., not with the round trip described by the residuals but with the time of flight (and so to the distance) between Earth and Moon.
Thus, the standard deviations reported in the figures of this work should be divided by two to obtain the relevant one-way standard deviation describing the precision in the measurements of the Earth-Moon distance. Following this prescription, the standard deviations for the one-way measurements of the 37, 32, and 69 points are 20 ps, 21.9 ps, and 21.5 ps, respectively, which, finally, correspond to a distance precision of 6 mm, 6.6 mm, and 6.5 mm. Correspondingly, we can compute the associated stability dividing by $\sqrt{N}$, where $N$ is the number of points. Thus, for the combined dataset of 69 points the stability is $6.5 \texttt{mm} / \sqrt{69} = 0.78$ mm. This means that with only 69 returns we effectively reach a sub-millimeter stability.

Furthermore, we pinpoint the second important difference with the analysis of the APOL NPs of Sect. \ref{APOLLOresults}: the standard deviation is reduced by $\sim 5$ orders of magnitude by the application of the detrending to the individual range measurements.
We also highlight that this millimeter precision is not limited by the capability of NGLR-1, but by the intrinsic capability of the hardware at the Wettzell station. This is very relevant, since an upgrade of the Wettzell station would guarantee a significant improvement in the precision of the measurement of the time of flight with NGLR-1 and the near-future next-generation retroreflectors to be deployed on the Moon.

\subsection{Results from the Grasse station}
\label{Grasseresults}

\begin{figure}
\centering
\includegraphics[width=\linewidth]{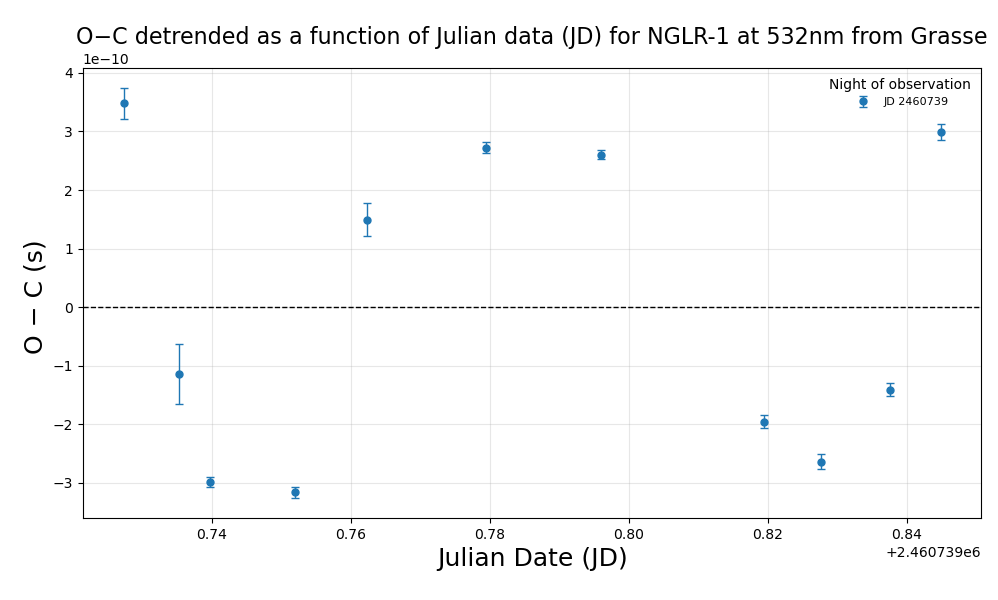}
\caption{Residuals O-C after the detrending for NGLR-1 data from the Grasse station at 532 nm.}
\label{fig:residuals_Grasse_532nm}
\end{figure}

For the Grasse data set, we follow the same analysis performed for the Wettzell data since we can work with the individual range measurements. However, in this case, we need to distinguish between the two subsets acquired at the different wavelengths of 532 nm and 1064 nm.\\
Starting from the 532 nm subset, after the statistical analyses described in Sect. \ref{detrending}, we choose a 3-degree polynomial detrending function. This is a conservative choice to avoid overfitting and a possible distortion of the true signal, but also a degree of 4 appears to be reasonable based on the results of the statistical analyses. In the following, to provide a comprehensive investigation, we also mention how our results would change if a 4-degree polynomial was preferred. Figure \ref{fig:residuals_Grasse_532nm} shows the residuals detrended with the 3-degree function, where the points are color-coded based on
the JD of observation, which, in this case, is the same one for all the measurements. As for the case of the Wettzell data presented in Figure \ref{fig:residuals_Wtzl}, the detrended residuals are randomly scattered around the zero line. This is a confirmation of our choice for the detrending function; indeed, we have checked that with a 2-degree polynomial function the O-C distribution still presents a pronounced oscillating trend with time.

\begin{figure}
\centering
\includegraphics[width=\linewidth]{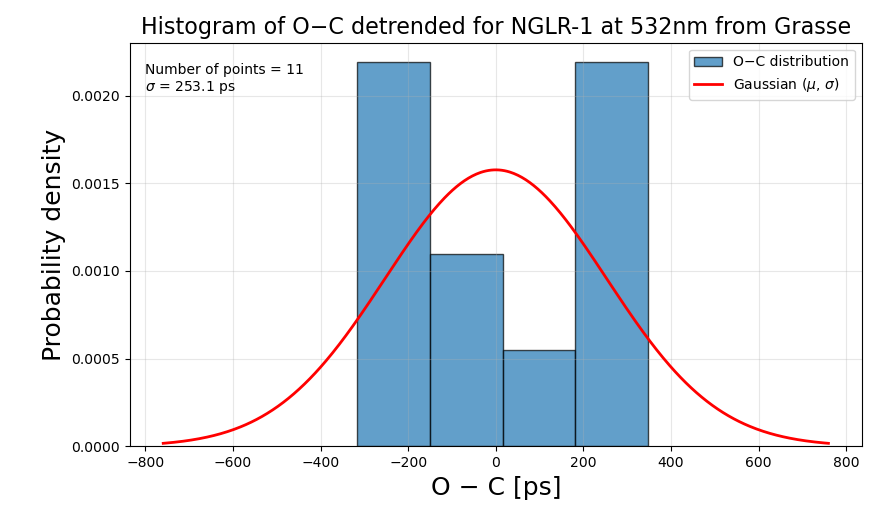}
\caption{Histogram of the residuals O-C after the detrending for NGLR-1 data from the Grasse station at 532 nm. The blue data illustrate the observations made by the Grasse station on the 4th of March 2025, while the red curve is the Gaussian curve with the standard deviation associated with the {same} data.}
\label{fig:hist_Grasse_532nm}
\end{figure}

Then, we study the distribution of these residuals to estimate the standard deviation. The result is provided in Figure \ref{fig:hist_Grasse_532nm}. {We here notice that the red Gaussian curve is not a fit to the blue data, but the Gaussian curve with the standard deviation associated to the data; this is the reason why the curve does not seem representative of the histogram, which does not present a central peak.}
Compared to the distribution of the whole Wettzell data set, the standard deviation is one order of magnitude larger, 253.1ps vs 43.0ps for the round trip (i.e. 126.6ps vs 21.5ps for the one-way range). This can be mainly ascribed to two factors: i) the Grasse observations at 532 nm are only 11 {points}, while the Wettzell ones are 37 and 32 {points} for the two subsets and 69 {points} for the combined data set, and ii) we have chosen the conservative 3-degree polynomial function for the detrending. Concerning the latter point, applying a 4-degree detrending function produces a distribution with $\sigma = 58$ps for the round trip, that is  $\sigma = 29$ps for the one-way. Thus, we expect that with increased statistics (i.e.{,} more observations at this wavelength) we would obtain a clearer hint to the proper polynomial degree to be applied and also a reduced standard deviation.

\begin{figure}
\centering
\includegraphics[width=\linewidth]{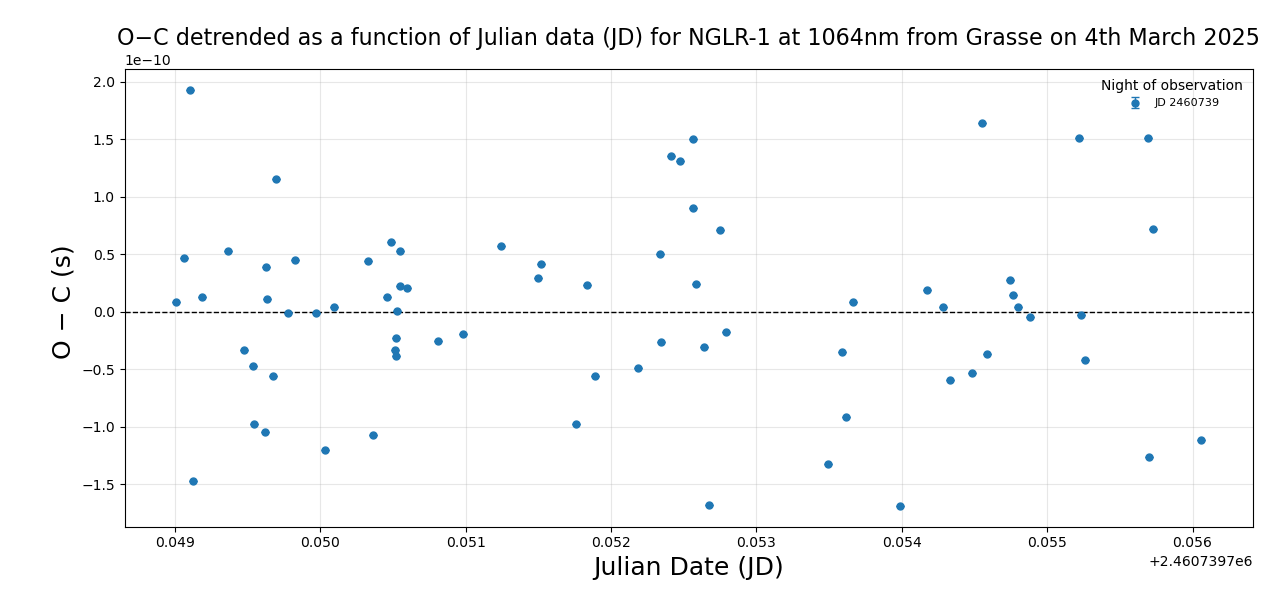}
\caption{Residuals O-C after the detrending for the 72 NGLR-1 individual range measurements from the Grasse station at 1064 nm on the 4th of March 2025.}
\label{fig:residuals_Grasse_1064nm_4march2025}
\end{figure}

\begin{figure}
\centering
\includegraphics[width=\linewidth]{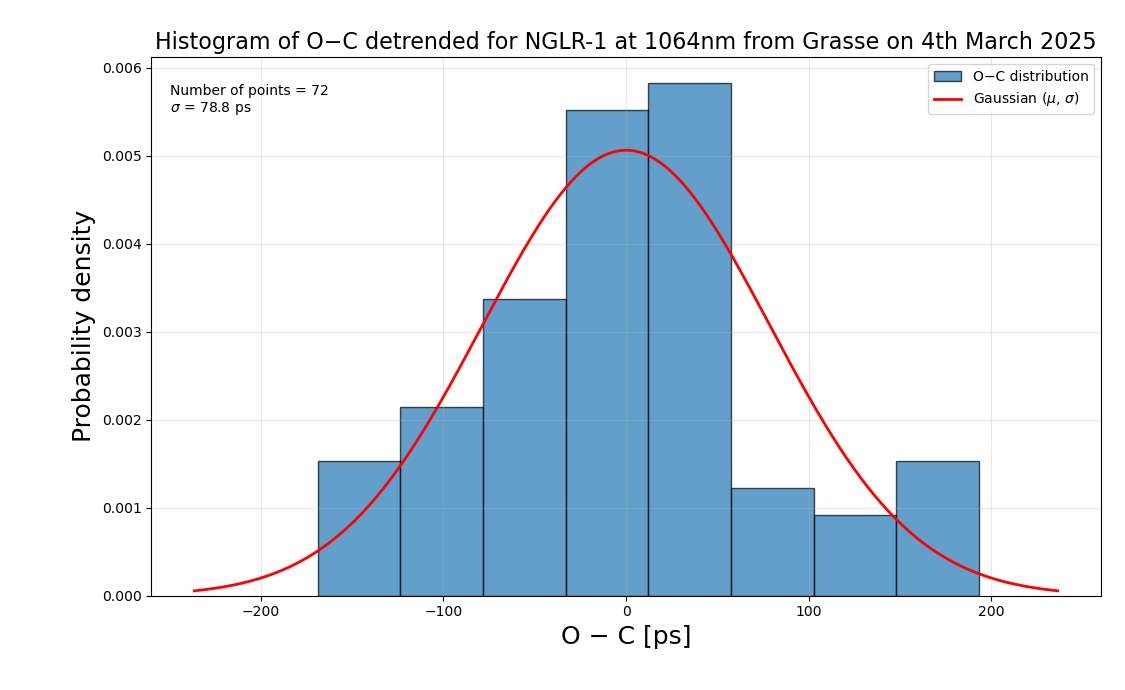}
\caption{Histogram of the residuals O-C after the detrending for the 72 NGLR-1 individual range measurements from the Grasse station at 1064 nm on the 4th of March 2025. The blue data illustrate the observations, while the red curve is the Gaussian curve with the standard deviation associated with the {same} data.}
\label{fig:hist_Grasse_1064nm_4march2025}
\end{figure}

To delve deeper into this prediction and to provide a more comprehensive study, we also analyze the Grasse data at 1064 nm.
The observations at this wavelength span a time period from the 3rd of March 2025 to the 30th of April 2025, so we need to select only a subset covering just one single day to apply the detrending procedure. For this reason, we choose the subset with {the most} individual range measurement observations, that is the one collected on the 4th of March 2025 in a time range between 17:58:29 and 18:08:45 UTC with 72 observations. The day of observation is also the same as the one of the Grasse data at 532 nm. This subset requires a 2-degree polynomial detrending that produces the residuals shown in Figure \ref{fig:residuals_Grasse_1064nm_4march2025} and the corresponding histogram provided in Figure \ref{fig:hist_Grasse_1064nm_4march2025}.

\begin{figure}
\centering
\includegraphics[width=\linewidth]{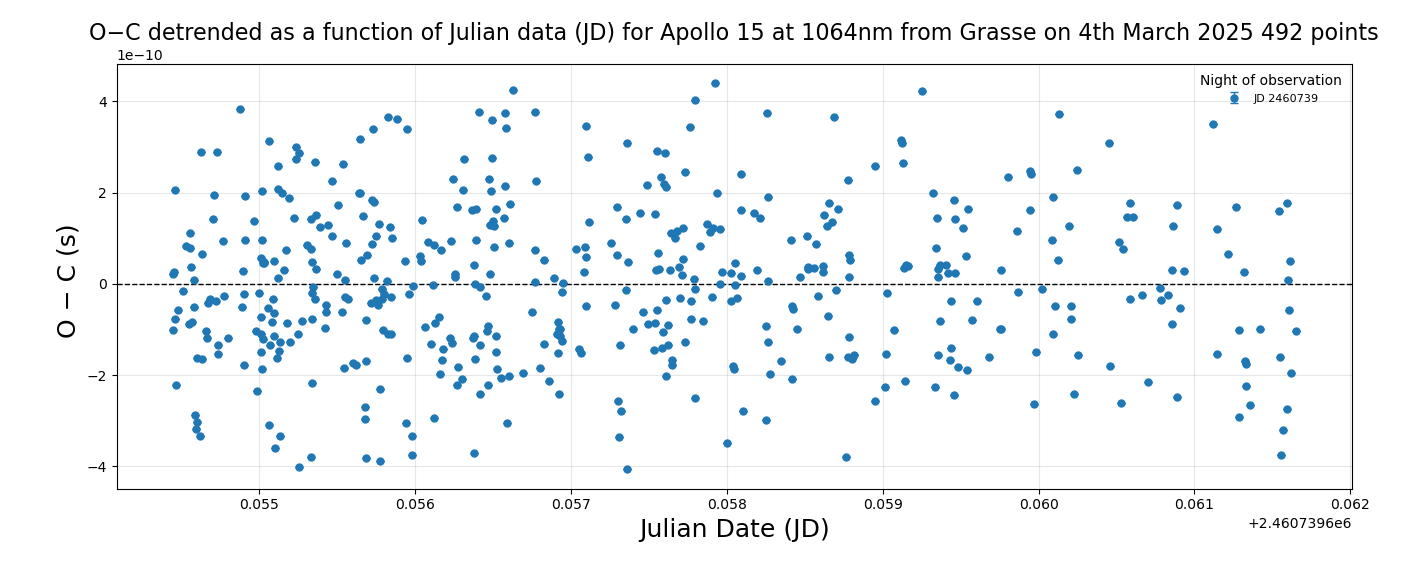}
\caption{Residuals O-C after the detrending for the Apollo 15 492 individual range measurements from the Grasse station at 1064 nm on the 4th of March 2025.}
\label{fig:residuals_Apollo15}
\end{figure}

\begin{figure}
\centering
\includegraphics[width=\linewidth]{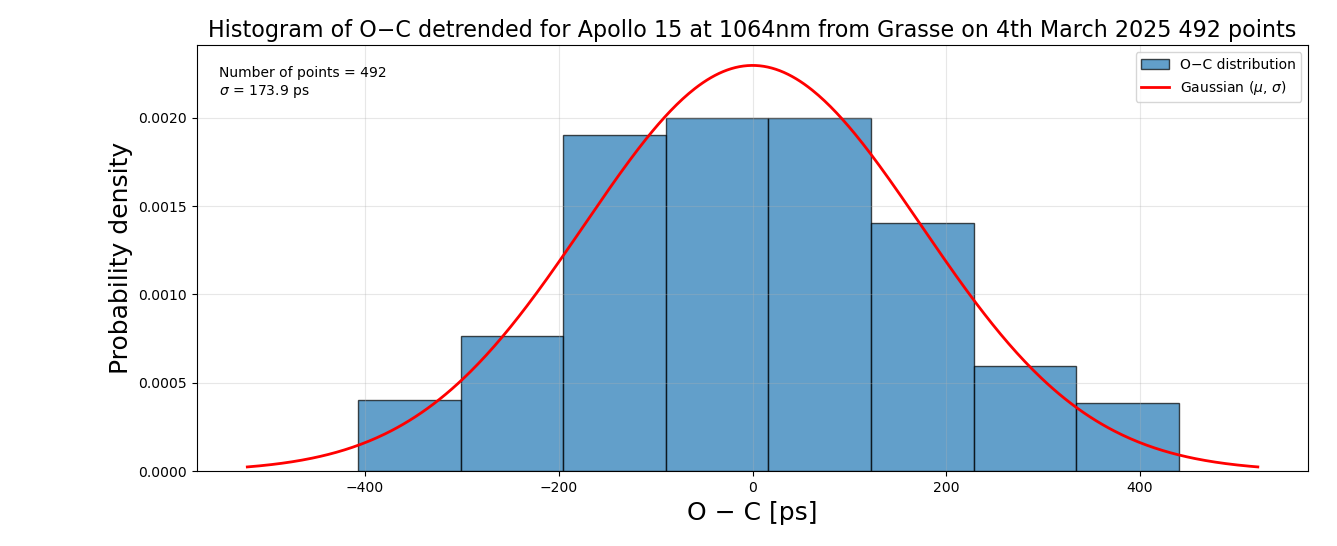}
\caption{Histogram of the residuals O-C after the detrending of the Apollo 15 492 individual range measurements from the Grasse station at 1064 nm on the 4th of March 2025. The blue data illustrate the observations, while the red curve is the Gaussian curve with the standard deviation associated with the {same} data.}
\label{fig:hist_Apollo15}
\end{figure}

This further inspection confirms that a data set with more points effectively provides a clear choice for the detrending function and, more importantly, an improved precision. Indeed, in this case, the standard deviation for the one-way measurement is $\sigma = 78.8/2 = 39.4$ps. As a consequence, this analysis is completely consistent with the one performed on the Wettzell data. Furthermore, similarly to the case of Wettzell, this precision is not limited by the capability of NGLR-1, but by the intrinsic capability of the hardware at the Grasse station \footnote{See Figure 3 at \url{https://www.physics.umd.edu/nglr/status_03122025.html} for reference.}.\\
To further support this result, we have also investigated the observations from Grasse to Apollo 15, which allow us to compare the precision of the same ground station on an "old-generation" retroreflector array, Apollo 15, and the first "next-generation" retroreflector, NGLR-1. More specifically, we have identified a subset of Apollo 15 data with substantial statistics: 492 individual range measurements acquired at 1064 nm on the 4th of March 2025 (that is also the same day analyzed for the other Grasse data sets) in a 10-minute range between 15:42:23 and 15:52:45 {UTC}. After a detrending with a 1-degree polynomial function, we have obtained the residuals and associated distribution shown in Figures \ref{fig:residuals_Apollo15} and \ref{fig:hist_Apollo15}, respectively.

Now, we can compare these results with the ones obtained from Grasse on NGLR-1 at the same wavelength and on the same {day of observations}. In particular, the 72 NGLR-1 points have provided a standard deviation for the one-way measurement that is $\sigma = 39.4$ps, while the 492 Apollo 15 points present a one-way standard deviation of $\sigma = 173.9/2 = 87.0$ps. Thus, it is clear that, considering the same ground station and the same day of observation, the precision of NGLR-1 is significantly improved compared to that of Apollo 15, {a bit over a factor of two better}, which strongly proves the power of the next-generation CCRs.
This comparison clearly {shows} the power of the single large-diameter retroreflector compared to the "old-generation" of retroreflector arrays, in this case Apollo 15.



\section{Future perspectives: the MoonLIGHT+MPAc payload}
\label{mpac}

The results presented in Sect.~\ref{results} strengthen the case for populating the lunar surface with additional next-generation, single large-diameter CCRs. Such deployment ensures that scheduled upgrades to terrestrial ground stations are fully exploited while sampling the Earth--Moon orbital geometry from several independent selenographic locations. The primary contribution of INFN-LNF to this ongoing effort is the MoonLIGHT+MPAc payload, an integrated instrument designed to provide sub-millimetre intrinsic ranging precision. A comprehensive description of the payload hardware, qualification program, and expected scientific return is given in \citet{muccino2025moonlight}.

\subsection{The MoonLIGHT CCR}
\label{moonlight}

MoonLIGHT (hereafter ML100) is a single, uncoated CCR manufactured from Suprasil 311, a high-purity, radiation-hard synthetic fused silica characterized by a near-zero coefficient of thermal expansion over a broad temperature range. This property drastically reduces internal thermal gradients that would otherwise induce optical wavefront distortions and degrade ranging efficiency during the lunar day. {In space-simulated tests with 14\,h of solar illumination, the 100\,mm Suprasil~311 CCR showed a maximum front-face temperature difference of only $\Delta T \approx 3$\,K ($2.9 \pm 1.4$\,K at $0^\circ$ and $3.0 \pm 1.4$\,K at $30^\circ$ incidence) and a thermal relaxation time $\tau_{\mathrm{CCR}} \approx (12\text{--}13) \times 10^{3}$\,s ($\approx 3.5$\,h) \citep{2017AdSpR..60.1300C}}. The CCR features a circular front-face clear aperture of 100\,mm, three nominally orthogonal reflecting faces with Dihedral Angle Offsets (DAOs) controlled within $0.2''$, and a front-face transmitted wavefront planarity better than $\lambda/10$ RMS, optimized for operation at $\lambda = 532$\,nm {\citep{muccino2025moonlight}}. 

The choice of a 100\,mm diameter balances return energy and diffraction spread. The angular radius of the central Airy disk of the Far Field Diffraction Pattern (FFDP) is $\theta \approx 1.22\,\lambda/d_{\rm ML} \approx 6.49\,\mu$rad, comfortably encompassing the apparent lunar velocity aberration vector ($4.12$--$4.55\,\mu$rad for operating LLR stations). Optical performance simulations carried out with CODE~V {optical design software}, combined with laboratory tests under simulated space environment at the INFN-LNF SCF\_Lab facility, demonstrate that the optical cross section remains stable across both linear polarization states. {The optical cross section averaged over the 4.0--4.5\,$\mu$rad velocity-aberration range ($\approx 0.75 \times 10^{9}$\,m$^{2}$ initially) showed only a transient reduction to $\approx 0.50 \times 10^{9}$\,m$^{2}$ after solar illumination, fully recovering within $\approx 5$\,h with no permanent degradation \citep{2017AdSpR..60.1300C}}. Furthermore, because return signal intensity scales with the fourth power of the aperture diameter ($I \propto d^4$), the pristine optical face of ML100 is predicted to deliver a return signal strength comparable to, or exceeding, the legacy Apollo 11 and 14 arrays, while completely eliminating the pulse-broadening libration tilt signature {\citep{muccino2025moonlight}}.

\subsection{The MoonLIGHT Pointing Actuator (MPAc)}
\label{mpachw}

In far-field conditions, the effective optical field of view of ML100 is confined to a conical envelope with an acceptance half-angle of approximately $17^\circ$. To guarantee maximum optical cross section, its front face must remain aligned within $3^\circ$ of the apparent Earth direction throughout the lunar synodic cycle. Because commercial lunar landers cannot guarantee such landing attitude tolerances on the lunar regolith, INFN-LNF developed the MoonLIGHT Pointing Actuator (MPAc) under an ESA development contract. The payload was selected by ESA and NASA for deployment at the Reiner Gamma magnetic swirl via a Commercial Lunar Payload Services (CLPS, CP-11) mission on board the Intuitive Machines Nova-C IM-3 lander, with launch expected in early 2027.

\begin{figure}
\centering
\begin{subfigure}[b]{.48\linewidth}
\includegraphics[width=\linewidth]{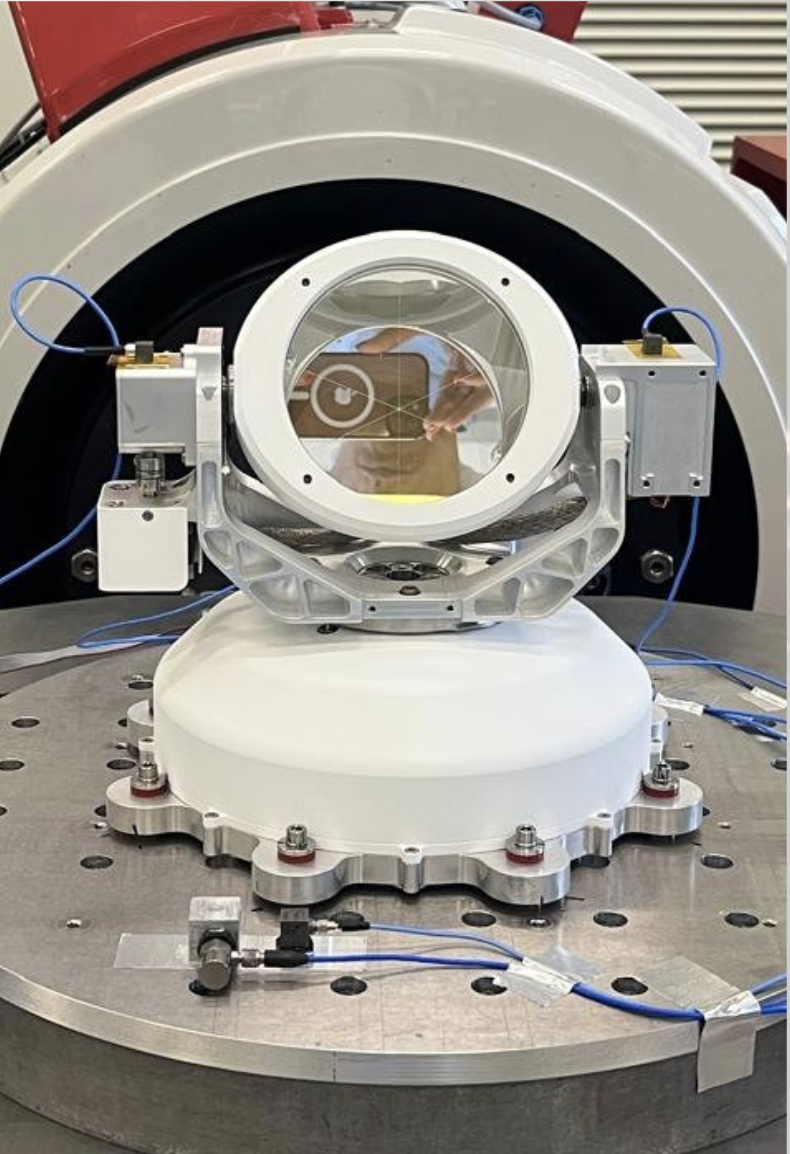}
\caption{Close-up view of MPAc: the 100 mm CCR and two-axis gimbal}
\end{subfigure}
\hfill
\begin{subfigure}[b]{.48\linewidth}
\includegraphics[height=9.0cm,angle=0,width=\linewidth]{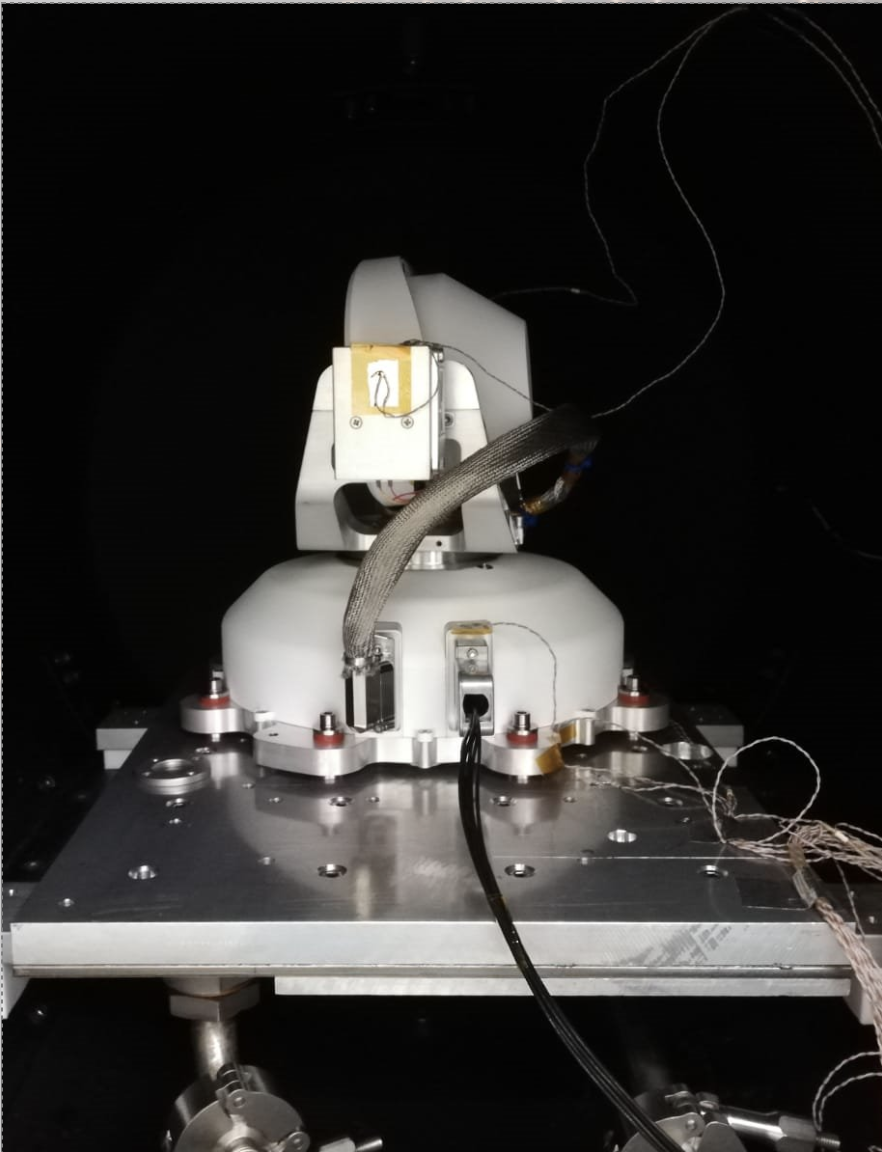}
\caption {MPAc during qualification test campaign at ESA-ESTEC}
\end{subfigure}
\caption{The Proto Flight Model 2 (PFM2) of MoonLIGHT+MPAc: the 100\,mm CCR is held within an elevation gimbal frame mounted on an azimuth baseplate. Reproduced from \citet{muccino2025moonlight} under CC BY 4.0.}
\label{fig:mpac_pfm1}
\end{figure}

MPAc (Figure~\ref{fig:mpac_pfm1}) is a robotic gimbal mechanism providing two independent rotational degrees of freedom. Its structural architecture consists of an azimuth frame interfacing the lander top deck, an elevation frame, and a thermally decoupled passive CCR housing. Stepper motors driving gearheads allow independent rotation from $0^\circ$ to $180^\circ$ in both axes, monitored by precision potentiometers and limit switches. The control electronics and flight software, hosted on an internal PCB ({Printed Circuit Board}), receive commanding from Earth via an RS-422 serial interface. After touchdown at the Reiner Gamma site ($7.585^\circ$\,N, $58.725^\circ$\,W), the lander orientation will be determined and the Nominal Earth Direction (NED) recalculated. MPAc will then execute an autonomous re-pointing maneuver, locking the ML100 normal along the center of the Earth libration pattern.
ESA contributed to the project also with specialized personnel who performed the thermal and structural analysis of MPAc.

\subsection{Expected impact on LLR analyses with PEP}
\label{mpacpep}

From an observational and analytical standpoint, MoonLIGHT shares with NGLR-1 the fundamental property of acting as an undispersed point reflector. Laser pulses reflected by ML100 preserve their incident temporal width, reducing the instrumental error budget to $\lesssim 1$\,mm RMS and shifting the dominant noise source entirely to the ground station timing systems and atmospheric calibration. In this context, facilities capable of transmitting and receiving multi-wavelength and polarization-controlled beams will be exceptionally well positioned to exploit ML100's optical characteristics.

\begin{figure}
\centering
\includegraphics[width=\linewidth]{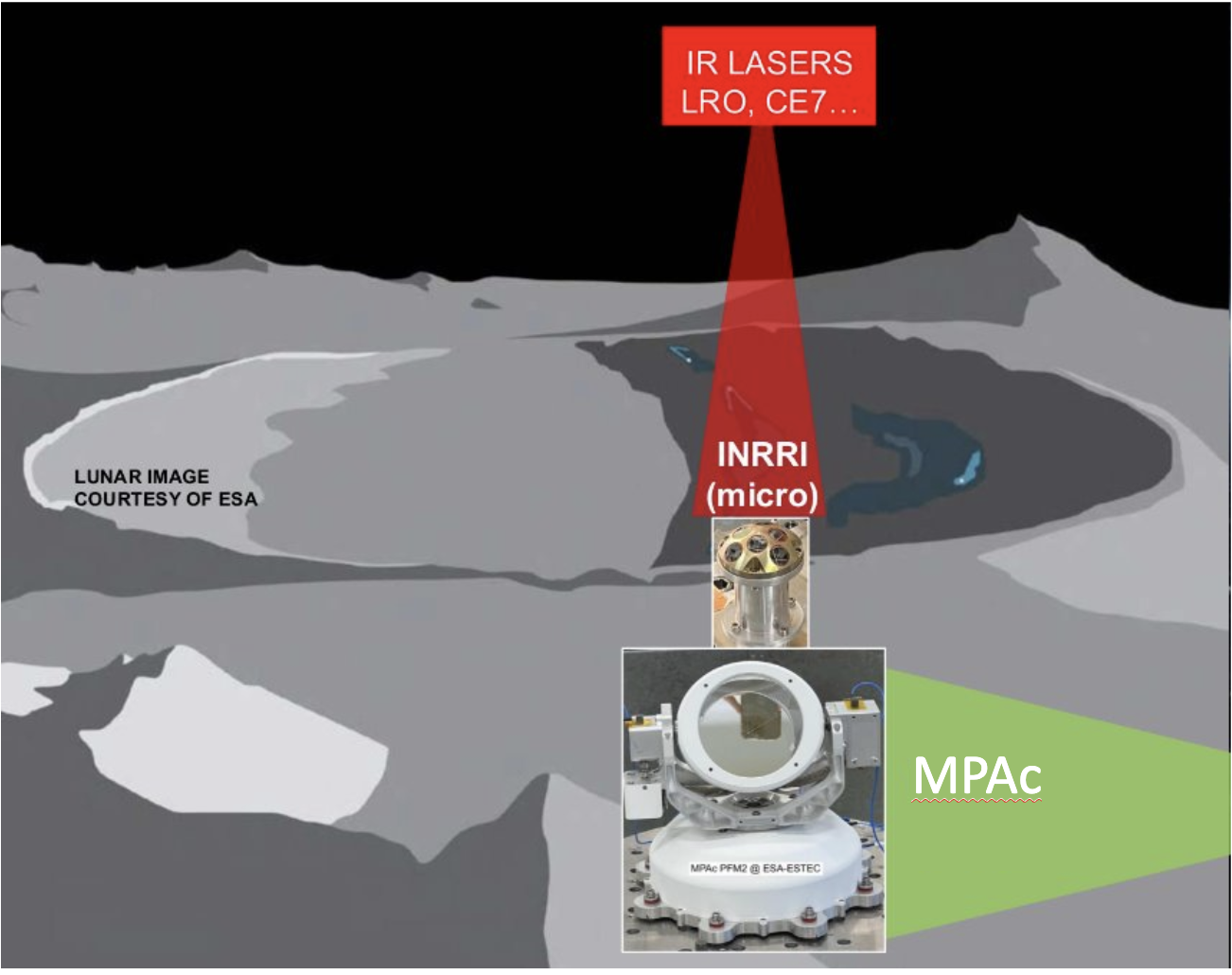}
\caption{Complementary observation geometries of the two laser
retroreflectors carried by IM-3. An infrared altimeter on a lunar
orbiter (Lunar Reconnaissance Orbiter (LRO), Chang'e-7) interrogates the INRRI microreflector along
the local vertical (red), while MoonLIGHT+MPAc is aimed at the Earth,
close to the local horizontal at this near-limb site (green).}
\label{fig:inrri_synergy}
\end{figure}

The analysis framework established in this study—incorporating CRD conversion to PEP format (Sect.~\ref{crdtomini}) and polynomial residual detrending (Sect.~\ref{detrending})—is directly applicable to MoonLIGHT observations upon landing. Incorporating high-cadence range measurements from multiple single-reflector stations across the lunar disk into PEP will break orbital-rotational parameter degeneracies. This is anticipated to enhance historical constraints on the weak and strong equivalence principles, the PPN parameters $\beta$ and $\gamma$, the time variability of the gravitational constant ($\dot{G}/G$), and geodetic precession by one to two orders of magnitude \citep{muccino2025moonlight}.
A further benefit arises at the IM-3 landing site from the combination
of Earth-based ranging and orbital laser altimetry
(Figure~\ref{fig:inrri_synergy}). Alongside MoonLIGHT+MPAc, the lander
carries as a commercial payload INRRI (INstrument for landing-Roving
laser Retroreflector Investigations), a passive microreflector of eight
fused-silica cube corner retroreflectors housed in an aluminium dome \citep{2017AdSpR..59..645D,2019SSRv..215....1P} as the one installed on Chang'e-6 \citep{2025SpScT...5..301W} and NASA's Perseverance rover (named LaRA, Laser Retroreflector Array, instead of INRRI). A range measurement
constrains the reflector position primarily along the
station--reflector direction, and only weakly in the two orthogonal
coordinates. Reiner Gamma lies at high selenographic longitude, close
to the western limb, so the Earth stands low above the local horizon
and the MoonLIGHT acceptance cone is directed close to the local
horizontal, whereas an infrared altimeter on board a lunar orbiter (such
as {Lunar Reconnaissance Orbiter} (LRO) or the upcoming Chang'e-7) interrogates INRRI along the local vertical. The
two fields of view are then nearly orthogonal, and each target
determines the coordinate to which the other is insensitive: the
MoonLIGHT coordinate transverse to the Earth beam is fixed by the
positioning of INRRI, and the INRRI coordinate transverse to the
orbital beam by the positioning of MoonLIGHT. The two payloads will
therefore jointly constrain the three-dimensional selenodetic position
of the deployment site --- and hence the station-to-reflector vector
entering PEP --- better than either can alone.

\section{Conclusions and future prospects}
\label{conclusions}

In this work, we have performed a detailed analysis of the Lunar Laser Ranging (LLR) data on NGLR-1 from the Earth stations of APOL, Wettzell, and Grasse in specific periods of observation, as described in Sect. \ref{data}. To this aim, we have used the open-source Planetary Ephemeris Program (PEP), specifically designed to process LLR data and compute their residuals through generated lunar ephemeris (see Sect. \ref{pep}). In this regard, we have outlined our methodology to implement the NGLR-1 data in PEP (Sect. \ref{crdtomini}) and to properly isolate and handle the true time of flight signal with the detrending procedure (Sect.\ref{detrending}). The main results of our analysis are reported in Sect. \ref{results} and summarized in the following:
\begin{itemize}
    \item The APOL data are released only in the format of normal points (NPs) without the individual range measurement data, so we have analyzed 46 NPs within $\sim$ 1 year of observation without the chance to apply the detrending. As a consequence, the NPs acquired during the same Julian {Day} (JD) show a trend with time (see Figure \ref{fig:residuals_APOLLO_withoutdetrending}) and the standard deviation of the residual distribution is large, of the order of $10^{-6}$ s, as reported in Figure \ref{fig:hist_APOLLO_withoutdetrending}. This is a useful study to investigate the behaviour of non-detrended NPs;
    \item The Wettzell data instead provide two NPs with, respectively, 37 and 32 individual range measurements, so, in this case, we perform the analysis with the detrending on both of these two data sets and the combined one. The results are as follows:
        \begin{itemize}
            \item the detrended residuals shown in Figure \ref{fig:residuals_Wtzl} lose their temporal trend in favour of a random distribution about the zero line and significantly reduce their order of magnitude, which becomes of the order of $\sim 10^{-11}$ s;
            \item the distributions of the detrended residuals visualized in Figure \ref{fig:hist_Wtzl} present a precision of the one-way Earth-Moon distance of about 6 mm (i.e. $\sigma \sim 21$ps in time).
        \end{itemize}
    \item The Grasse data have been analyzed for three specific subsets, which are NGLR-1 at 532 nm and 1064 nm and Apollo 15 at 1064 nm, providing the following results:
        \begin{itemize}
            \item the small data set of 11 points from NGLR-1 at 532 nm presents a large standard deviation due to an unclear choice for the detrending function and the small number of individual range measurements. This conclusion is also corroborated by the following point;
            \item the larger data set of 72 points from NGLR-1 at 1064 nm acquired in a specific 10-minute range on the 4th of March 2025 presents a clear 2-degree polynomial detrending instead and provides a one-way standard deviation of 39.4ps (see Figure \ref{fig:hist_Grasse_1064nm_4march2025}), significantly improved compared to that at 532 nm and consistent with the one obtained with the Wettzell data set;
            \item the big data set of 492 points from Apollo 15 at 1064 nm acquired in a specific 10-minute range on the 4th of March 2025, after a 1-degree detrending, provides a one-way standard deviation of $\sigma = 87.0$ps, as shown in Figure \ref{fig:hist_Apollo15}, clearly {showing} the supremacy of next-generation CCRs on retroreflector arrays.
        \end{itemize}
\end{itemize}
Finally, the results of this work are robust evidence of the powerful contribution of NGLR-1, and of future next-generation lunar retroreflectors, in view of increasing the precision in the measurement of the distance between the Earth and the Moon. Currently, the precision provided by the Wettzell and Grasse stations to NGLR-1 is not limited by the specifics and the optics of the retroreflector itself, but by the hardware capabilities of the Earth stations. This implies that an upgrade of the Earth stations would enable a further improvement in the LLR measurements, pushing the single large-diameter retroreflectors towards their effective potential. In this regard, the Wettzell hardware is close to the limit of what can be achieved to date in terms of the precision, while the planned upgrades rather focus on an increased data yield, to enhance the statistics and/or improve the stability. \\
Assuming also a parallel improvement of LLR analysis software packages (like PEP and others) in the modeling of the Moon, this scenario of high-precision LLR observations paves the way for improved tests of General Relativity (GR) and gravity theories beyond GR{, such as spacetime torsion, f(R) gravity, and non-minimally coupled gravity}. In this framework, the next INFN-LNF contribution to the LLR space segment is the MoonLIGHT+MPAc payload described in Sect. \ref{mpac}, a single 100 mm CCR mounted on a double-axis pointing actuator that will align its front face towards the Earth after landing at the Reiner Gamma swirl with the NASA CLPS/CP-11 mission \citep{muccino2025moonlight}; the analysis pipeline validated here on NGLR-1 is directly applicable to its future observations.

\section*{Acknowledgments}
The INFN authors gratefully acknowledge the overall guidance 
and support of Istituto Nazionale di Fisica Nucleare (INFN), as part
of the MoonLIGHT-2 experiment in the framework of the research
activities of the Commissione Scientifica Nazionale No. 2 (CSN2). The INFN authors also wish to acknowledge the precious and skilled technical support of Mattia Tibuzzi and Lorenzo Salvatori (of INFN-LNF) on the mechanical design, manufacturing, assembly and integration of MPAc.
We warmly thank also ESA-INFN Contract n. 4000133721/21/NL/CR
(INFN code TTA\_20LNF\_140) for the support to MPAc and the flight
opportunity that ESA has secured by means of a high level, legally binding
NASA-ESA Memorandum of Understanding. INFN authors also recognize
the support from the Agreement ASI-INFN n. 2019-15-HH.0 on
"Sistemi Avanzati Laser per Applicazioni Spaziali: Attività Congiunte
presso INFN-LNF", which was terminated by ASI on April 1, 2026,
before the completion of this paper and about 1 year before the launch to the Moon of
MoonLIGHT+MPAc. We are grateful to N. Colmenares for the fruitful
collaboration on the conversion of the LLR data and for the discussion
about the APOL normal points, to C. Courde and J. Chabé for sharing Grasse data on NLGR-1,
and to J. Eckl for the release of the Wettzell data on NLGR-1. D. Currie wishes to acknowledge the NASA funding received at the University of Maryland from NASA.


\section*{Data/Software/Code Availability}

The data sets used in this work for the LLR observations from APOL and Wettzell stations are publicly available, while those from Grasse have been privately shared by C. Courde.
In addition, the software PEP is publicly available at \url{https://gitlab.com/jbattat/} and the codes used for the analyses reported in this study will be shared upon reasonable request to the corresponding author.



\bibliographystyle{mnras}
\bibliography{bibliografia} 




\appendix



\label{lastpage}
\end{document}